\documentclass[12pt]{iopart}

\usepackage{graphicx}
\usepackage{amssymb}
\usepackage{color}

\begin{document}

\title[Suppressed Density of States in Self-Generated Coulomb Glasses]{Suppressed Density of States in Self-Generated Coulomb Glasses} 

\author{Louk~Rademaker$^{1,2}$,
Zohar~Nussinov$^3$, 
Leon~Balents$^{1}$ and Vladimir Dobrosavljevi\'{c}$^4$}

\address{$^1$ Kavli Institute for Theoretical Physics, University of California Santa Barbara, California 93106, USA}
\address{$^2$ Perimeter Institute for Theoretical Physics, Waterloo, Ontario N2L 2Y5, Canada}
\address{$^3$ Physics Department, CB 1105, Washington University, 1 Brookings Drive, St. Louis, MO 63130-4899}
\address{$^4$ Department of Physics and National High Magnetic Field Laboratory, Florida State University, Tallahassee, Florida 32306, USA}

\begin{abstract}
{
We investigate the structure of metastable states in self-generated Coulomb glasses. In dramatic contrast to disordered electron glasses, we find that these states lack marginal stability. Such absence of marginal stability is reflected by the suppression of the single-particle density of states into an exponentially soft gap of the form $g(\epsilon) \sim e^{-V / \xi |\epsilon|}$. 
To analytically explain this behavior, we extend the stability criterion of Efros and Shklovskii to incorporate local charge correlations, in qualitative agreement with our numerical findings.
Our work suggests the existence of a new class of self-generated glasses dominated by strong geometric frustration.
}
\end{abstract}

\noindent{\it Keywords\/}: Self-generated glasses, Efros-Shklovskii gap, marginal stability

\submitto{\NJP}

\maketitle

%
%

%
%
%
%



How and why glasses form is one of the most intriguing open questions of modern science. A glass is a rigid material yet lacks crystalline order. It has been recognized thousands of years ago that glasses can be made by fast-cooling a liquid below its solidus temperature to avoid crystallization. The resulting supercooled liquid displays an exponential increase of its viscosity. As such, glasses are inherently non-equilibrium states of matter, even though on any reasonable experimental timescale the material appears to be static.\cite{Berthier:2011hs,Debenedetti:2001bh,2016RPPh...79a6601K,Langer:2014ic}

The natural question is why the crystal does not nucleate inside the supercooled liquid phase. One of the possible answers revolves around the existence of local short-range order in both the liquid, the supercooled liquid and the glass phase. If the short-ranged density correlations in the liquid are manifestly different than those characterizing the preferred crystalline state, a glass can be formed through fast-cooling. The system then freezes into a locally ordered frustrated configuration that requires a macroscopic number of rearrangements to lower the energy and realize the standard crystalline form.

An example is the icosahedral short-range order first proposed\cite{1952RSPSA.215...43F} and later observed in metallic glasses such as Ti-Zr-Ni.\cite{2003PhRvL..90s5504K,2009PhRvL.102e7801S,2013PhRvB..87r4203S} A single icosahedron is much denser packed than any crystalline structure like fcc or hcp, which suggests that in the liquid phase the dominant density correlations are icosahedral. However, one cannot fill space with icosahedral order, leading to glassy behavior at low temperatures.

A more recent example is found in self-generated - which means no quenched disorder - electron glasses.\cite{Kagawa:2013hz,Sato:2014jp,Sato:2014cl,Kagawa:2017de} In the organic layered materials of the $\theta$-family, the electrons display distinct glassy features after fast-cooling to avoid a stripe ordering transition. Upon cooling, the local charge order present in the high-temperature liquid strengthens even further within the glassy regime.

\begin{figure}
  \includegraphics[height=6cm]{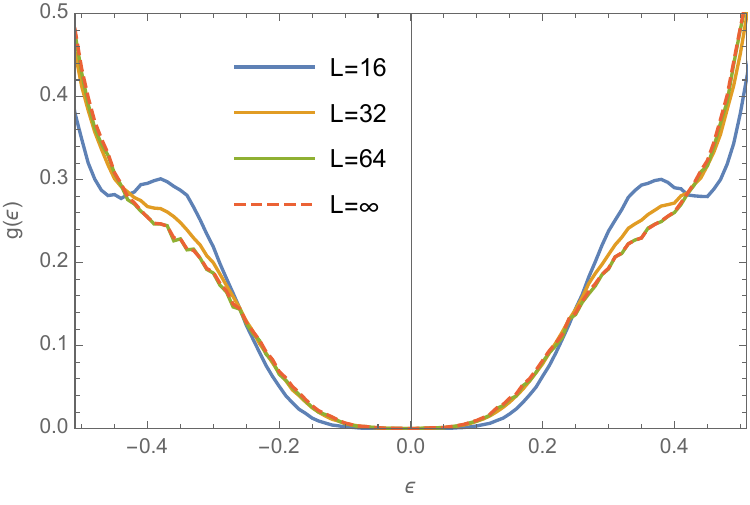}
  \includegraphics[height=6cm]{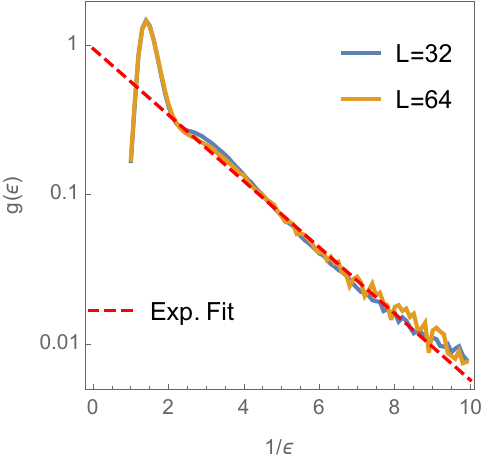}
  \caption{The exponentially suppressed Coulomb gap for our model of a self-generated glass on a triangular lattice. \\
  {\bf Left:} The density of states $g(\epsilon)$ as a function of energy $\epsilon$ around the Fermi level $\epsilon = 0$, averaged over an ensemble of metastable particle configurations. The density of states is normalized so that $\int g(\epsilon) d\epsilon = 1$, and energy is expressed in units of the nearest neighbor repulsion $V$. For $L=16$ we average over $N_{ms} = 50000$ metastable states, for $L=32$ over $N_{ms} = 16336$ and for $L=64$ we have $N_{ms} = 424$ metastable states. The dashed line is the extrapolation for $L=\infty$ assuming that $g(\epsilon, L) = g(\epsilon, \infty) + \mathrm{const.} \times L^{-2}$. The bump around $|\epsilon|\sim 0.3$ for small $L$ is an irrelevant finite size effect. \\
  {\bf Right:} To show that we find an exponentially suppressed density of states near the Fermi level, we show an Arrhenius plot of the density of states - which means the logarithm of $g(\epsilon)$ versus $1/|\epsilon|$. We find a clear regime where the density of states follows $g(\epsilon) = a e^{-b/|\epsilon|}$, and the dashed line shows the fit of this form. Note that quenched disorder Coulomb glasses are expected to have a linear density of states, $g(\epsilon) \sim \epsilon$, which is clearly violated here. }
  \label{FigFit}
\end{figure}

In this Letter we propose, based on the assumption of glassiness due to locally frustrated order, that such self-generated glasses display a characteristic soft gap in the single-particle density of states. This implies that self-generated glasses are \emph{not} marginally stable,\cite{Muller:2015kl} which would require saturation of the Efros-Shklovskii bound\cite{1975JPhC....8L..49E,1976JPhC....9.2021E}. We observe that the presence of local charge order stabilizes the glass and suppresses the density of states to $g(\epsilon) \sim e^{-V/\xi |\epsilon|}$ where $V$ is the interaction strength, and $\xi$ is the finite correlation length of the local order. As can be seen in Fig. \ref{FigFit}, this form is consistent with numerical simulations.

To derive this, we provide analytical arguments based on the Efros-Shklovskii stability criterion, which are then further constrained by the local charge order. We numerically verify our claim using a model of long-range interacting spinless electrons on a half-filled triangular lattice. This model was introduced earlier to describe the glassy behavior of electrons in the $\theta$-organic compounds.\cite{Mahmoudian:2014vh,Rademaker:2016fj} We find a remarkable agreement between our analytical form of the density of states and the numerical results. Finally, we address the limitations and implications of our model, and the relation to models with quenched disorder.

\section{Definitions}
We consider a model system of particles on a triangular lattice. On each lattice site the density is given by $n_i = 0,1$, and the total energy is given by the Coulomb repulsion between the particles,
\begin{equation}
	E = \frac{1}{2} \sum_{i,j \neq i} \frac{V}{|\vec{r}_{ij}|} (n_i - \overline{n}) (n_j - \overline{n}),
	\label{Energy}
\end{equation}
where $\overline{n}$ is the average number of particles per site, $\vec{r}_{ij}$ the (dimensionless) distance between site $i$ and $j$ and $V$ some unit of energy. We define the on-site energies as $\epsilon_i = \sum_{j \neq i} \frac{V}{|\vec{r}_{ij}|} (n_j - \overline{n}) $ such that the total energy is $E = \frac{1}{2} \sum_i \epsilon_i (n_i - \overline{n}) $. It is important to note that we consider self-generated glasses only with a translationally invariant Hamiltonian, in contrast to the traditional electron glasses that require quenched disorder\cite{Pollak:2013vs}. Details on the glassy behavior of this model are presented in Ref. \cite{Mahmoudian:2014vh}, where we also showed that the effective classical description is still valid upon including a small quantum hopping term.

In the glass at low temperatures we assume the particles are frozen into a stable nonperiodic configuration. The stability requirement implies that if we move a particle from site $i$ to site $j$, the total energy of the system should increase,\cite{1975JPhC....8L..49E}
\begin{equation}
	\Delta E = \epsilon_j - \epsilon_i - \frac{V}{|\vec{r}_{ij}|} > 0.
	\label{Stability}
\end{equation}
The ground state of the system naturally satisfies this single-particle stability criterion. Any configuration that satisfies the stability criterion is called a \emph{metastable state}. Note that these states are sometimes called 'inherent structures' to avoid confusion with the thermodynamic notion of metastability, or 'pseudo-ground states'. Also note that this definition is not unique.: one can define metastability with respect to one- \'{a}nd two-particle hops; one, two \'{a}nd three-particle hops; or local single-particle hops only, etcetera. However, in this paper we stick to the definition following Eqn. (\ref{Stability}). 

Following the arguments mentioned in the introduction, we assume that these metastable states have some kind of local short-range order. This is characterized by the density correlation function
\begin{equation}
	\Pi_{ij} = \langle (n_i - \overline{n}) (n_j - \overline{n}) \rangle
	\label{DensDensDef}
\end{equation}
where the average $\langle \cdots \rangle$ is over the ensemble $\Gamma_{MS}$ of metastable states. In typical systems without long-range order, the density correlations decay exponentially at large distances while at shorter distances a certain modulation is expected. We therefore propose to approximate the density correlation function by
\begin{equation}
	\Pi(\vec{r}) \sim
		\sum_{\vec{M}} \cos  \left( \vec{M} \cdot \vec{r} \right)
		e^{-r / \xi_M},
\end{equation}
where the sum runs over the short-range density modulation wavevectors $\vec{M}$, and $\xi_M$ is the correlation length for that given wavevector. The Fourier transform to momentum space yields a sum over Lorentzians,
\begin{equation}
	\Pi(\vec{k}) = \sum_{\vec{M}} \frac{A_M}{(\vec{k} - \vec{M})^2 + \xi^{-2}_M},
	\label{LorentzianShape}
\end{equation}
which we will assume is the generic form of the density correlation function in the remainder of this paper.

\section{Numerical studies}

We start by numerically studying the ensemble of metastable states for the model of particles on a half-filled ($\overline{n} = 1/2$) triangular lattice of size $N = L^2$ with periodic boundary conditions, introduced in Ref. \cite{Mahmoudian:2014vh,Rademaker:2016fj} and in Eqn. (\ref{Energy}). The interaction potential contains both a Coulomb tail and nearest-neighbor repulsion, tunable by the parameter $x$,
\begin{equation}
	V(\vec{r}) = x \frac{V}{|\vec{r}_{ij}|} + (1-x) V \delta_{|\vec{r}_{ij}|=1}.
	\label{Potential}
\end{equation}
The short-range component of the interactions reflects the fact that in realistic sytems, such as the organics, the charge distribution on a single site is not exactly point-like. At longer distances, the nontrivial shape of the onsite orbitals can be neglected and we regain the standard Coulomb term.

The long-range nature of the Coulomb interaction is taken care of using Ewald summation on a parallelogram-shaped super unit cell with periodic boundary conditions.\cite{1996CoPhC..95...73T} When only nearest neighbor interactions in the second term in Eqn.~(\ref{Potential}) are present, there is an exponentially large number of degenerate ground states that are not separated by barriers. The inclusion of a Coulomb tail lifts the degeneracy and creates barriers between different configurations. 

To obtain an ensemble of unique metastable states numerically, we start with a completely random configuration and lower the energy by random single-particle moves (including nonlocal moves) until the stability criterion is explicitly met. This means that we checked all the possible short- and long-range moves to make sure none of such moves can increase the energy. 

Once we have a metastable configuration, we check whether it is a unique configuration. We explicitly check uniqueness by comparing the newfound configuration to all the earlier found metastable states, including all $6L^2$ possible rotations and translations. The result for the counting of the number of metastable states is shown in Table \ref{TableNMS}. We have for a given system size $L$ a collection of $M$ random initial configurations, from which a stable configuration is constructed as described above. This way we find a set of $N_{ms}(M)$ unique metastable states.

Note that we explicitly refrain from using any approximate procedures that would allow us to study bigger systems.\cite{Mobius1992} At the one hand, we already find sufficient convergence of results to make claims about the functional form of the low-energy density of states. On the other hand, the use of an approximate procedure without explicitly checking the stability criterion prevents us from correctly counting the number of metastable states.

For $L=4$ we can explicitly check all possible configurations, and we find $N_{ms}=3$ unique metastable configurations. For $L=6$, we see that the number of unique metastable configurations saturates to a value of $N_{ms}=93$. The approach to this saturation value for smaller values of $M$ gives us a function $N_{ms,6}(M)$. For larger systems $L>6$ we approximate the expected total number of metastable states by scaling $\alpha^{-1} N_{ms,6}(\alpha M)$. This works for $L=8$ and $L=10$. However, for $L=12$ this estimate is conservative and provides a very weak lower bound on the total number of metastable states.

\begin{table}

\begin{tabular}{llllll}
$L$ & $2^{L^2}$ & $M$ & $N_{\mathrm{ms}} (M) $ & expected total $N_{\mathrm{ms}} $ & Complexity $S =L^{-2} \log N_{ms}$\\
\hline
4 & 65536 & 65536 & 3 & 3 &  $0.069$ \\
6 & $7 \times 10^{10}$ & 50000 & 93 & 93 & $0.126$  \\
8 & $2 \times 10^{19}$ & 50000 & 13093 & $\sim$ 17500 & $\sim 0.153$\\
10 & $10^{30}$ & 31950 & 31462 & $\gtrsim 8 \times 10^6$ & $ \gtrsim 0.159$ \\
12 & $2 \times 10^{43}$ & 21811 & 21810 & $ > 10^9$ &  $> 0.14$ \\
\end{tabular}

\caption{\label{TabStates} The number of distinct metastable states after $M$ sweeps for the lattice sizes $L=4$ to $L=12$. The expected total number of unique metastable states is obtained by scaling the curves $N_{ms}(M)$ to match the result for $L=6$. The estimate for $L=12$ is extremely conservative, and is possible larger.}
\label{TableNMS}
\end{table}

\begin{figure}
  \includegraphics[width=\columnwidth]{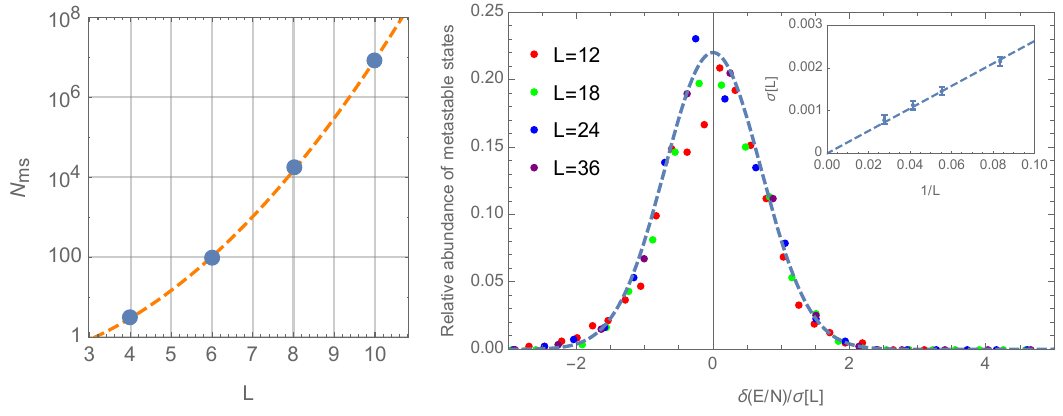}
  \caption{
  Left: A logarithmic plot of the number of metastable $N_{ms}$ states versus linear lattice size $L$, for the model of Eqn. (\ref{Potential}) with $x=1$. The fit $N_{ms} \sim 2^{0.25 L^2}$ indicates an exponential number of metastable states.  Right: The distribution of the energy density of the unique metastable states, for $L=12$, 18, 24, and 36. The distribution is fitted with a Gaussian around the same average energy $\overline{E}/N = -0.167$ (which is higher than the striped ground state energy density \cite{Mahmoudian:2014vh}) and width $\sigma(L)$. The standard deviation $\sigma(L)$ decreases with increasing system size, as shown in the inset, suggesting that in the thermodynamic limit there are infinitely many metastable states with the same energy density.}
  \label{metastates}
\end{figure}

For these larger system sizes we can thus compute the complexity, defined as the entropy associated with the number of metastable states, $S = \frac{1}{L^2} \log N_{ms}$. If there are less than exponentially many metastable states, the complexity should vanish. The results suggest that this is not the case, and that the complexity approaches a value $0.14-0.16$ in the thermodynamic limit, see also Fig. \ref{metastates}. This scaling is consistent with typical models of glass formation that exhibit an exponential number of metastable states, such as the random Sherrington-Kirkpatrick model \cite{Debenedetti:2001bh,Bray:1980cy,Monasson:1995dw,Cavagna:2004gm}.

\begin{figure}
  \includegraphics[width=\columnwidth]{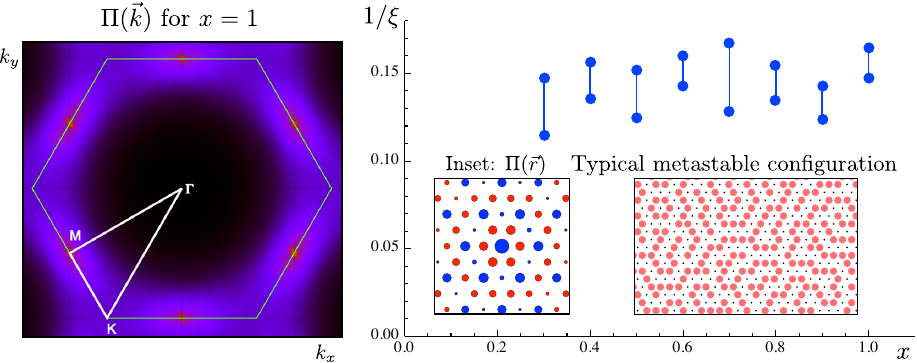}
  \caption{Left: The density correlation function $\Pi(\vec{k})$ for $x=1$ averaged over an ensemble of 996 metastable states on a $N=48 \times 48$ lattice. Clear peaks at the $\vec{M}$-points are seen, indicating local charge order. Right: The extracted inverse correlation length at the $\vec{M}$-point as a function of $x$ by fitting $\Pi(\vec{k})$ to a Lorentzian shape of Eqn. (\ref{LorentzianShape}). The upper and lower bounds indicate the error bar. Right, inset: Real space density correlations $\Pi(\vec{r})$ at $x=1$. The size of the balls indicate the strength of $\Pi(\vec{r})$ whereas the blue (red) indicates positive (negative) correlations. We also include a snapshot of a typical metastable configuration.}
  \label{FigDensCorr}
\end{figure}

The ensemble of metastable states has a narrow distribution of energies. In the thermodynamic limit, all metastable states have the same energy density $\overline{E}/N$, see Fig. \ref{metastates}, right. Consistent with our assumptions, the ensemble has local charge correlations at the  $\vec{M}$-points of the hexagonal Brillouin zone, see Fig. \ref{FigDensCorr}, left. From a depiction of this charge order in real space, including a typical snapshot of a metastable state, it is clear that there exists no periodic arrangement of discrete particles that has this local structure, see Fig. \ref{FigDensCorr}, right inset. As shown in the same figure, the correlation length $\xi$ we extract from $\Pi(\vec{k})$ by fitting to Eqn. (\ref{LorentzianShape}) seems to be independent of the value of $x$.

We next proceed to compute the density of states (DOS). This is obtained by making a histogram of the on-site energies, averaged over a large ensemble of metastable configurations. For the pure Coulomb interaction ($x=1$), we computed the DOS for linear lattice size $L=16, 32$ and $64$, and extrapolated these DOSs to $L=\infty$. The results are shown in Figure \ref{FigFit}. Though we see that the DOS rapidly converges with increasing system size, we avoid making conclusions for the energy regime below $|\epsilon|~0.1$ due to the scarcity of data points. The best fit for the density of states $g(\epsilon)$ at low energies, but still above $\epsilon ~ 0.1$, is given by the functional form $g(\epsilon) = a e^{-b/|\epsilon|}$, rather than the $g \sim |\epsilon|$ powerlaw form expected from the Efros-Shklovskii bound.

\begin{figure}
  \includegraphics[width=0.6\columnwidth]{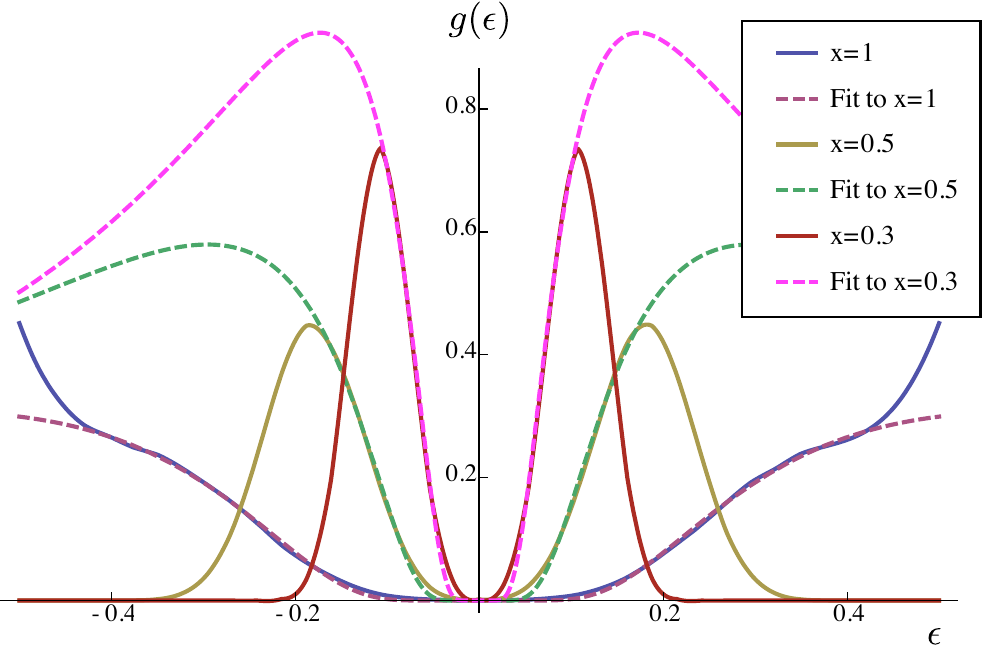} \\
  \includegraphics[width=0.6\columnwidth]{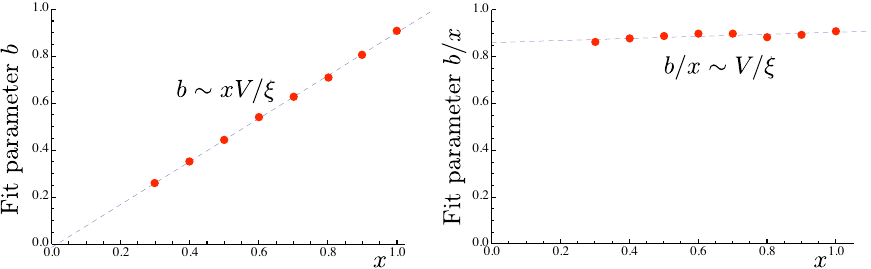}
  \caption{The density of states $g(\epsilon)$ for various values of $x$ given the interaction potential Eqn. (\ref{Potential}), for $L=48$ and an ensemble of 1000 metastable states. We fit the numerical density of states at low energies with the functional shape $g(\epsilon) \sim e^{-b/|\epsilon|}$, as shown in the top panel. The fit parameter $b$ as a function of $x$ is shown in the lower panel, consistent with the central result of Eqn. (\ref{BasicRelation}) we find $b \sim xV/\xi$. Also note that the typical temperature of a glass is about 0.01 $V$, well within the hardest part of the gap.}
  \label{FigTotalDos}
\end{figure}

For other values of $x<1$, we have a smaller set of metastable states up lattices of linear size $L=48$ to obtain the density of states, see Fig.\ref{FigTotalDos}. We fit again the low-energy density of states by an exponential form $g(\epsilon) = a e^{-b/|\epsilon|}$. The parameter $b$ as a function of $x$ is shown in the lower panel of Fig.\ref{FigTotalDos}, where we find that $b$ is proportional to $x$. 

In conclusion, we found unambiguously that the low energy density of states in the self-generated Coulomb glass is suppressed with respect to the Efros-Shklovskii powerlaw gap. The degree of exponential suppression is directly proportional to the strength of the Coulomb interactions. In the following section, we aim to understand this suppression qualitatively. 


\section{Analysis of the Coulomb Gap}

The spectrum of a long-ranged ordered state would have a hard gap in the spectrum. This gap is now smeared due to the spatial charge fluctuations reflecting the amorphous nature of the metastable states. In this section we will present analytical arguments how these charge fluctuations can qualitatively cause the resulting exponential gap. 

Recall that the traditional Efros-Shklovskii arguments\cite{1975JPhC....8L..49E,1976JPhC....9.2021E} use the stability criterion Eqn. (\ref{Stability}) to restrict the possible position of low-energy particles. We will use the presence of charge correlations to further constrain these positions.

For a given metastable state in the ensemble $\Gamma_{MS}$, we define the origin as some empty site with onsite energy $\epsilon_0$ asymptotically close to zero. The probability that a site at the position $\vec{r}$ is occupied equals
\begin{equation}
	P_e (\vec{r}) = \frac{1}{2} - 2 \Pi (\vec{r}).
	\label{Prob1}
\end{equation}
Because of the local order, the density correlation function $\Pi(\vec{r})$ is the product of an oscillating function and an exponentially decaying function. Since the precise wave-vector of the local order is irrelevant to further considerations, we will only consider distances $\vec{r}$ that are commensurate with the wavelength of the order, that is $\vec{M} \cdot \vec{r}$ is a multiple of $2\pi$; as we will see, considerations for these distances will lead to strong constraints. We will denote the density correlation function at such commensurate sites by $\widetilde{\Pi}(\vec{r}) \sim \frac{1}{\sqrt{r}} e^{-r / \xi}$.

Next we introduce the local distribution of on-site energies $g_{\vec{r}} (\epsilon)$ at distances $\vec{r}$. Since an occupied site has negative onsite energy, the probability to find a particle at position $\vec{r}$ equals
\begin{equation}
	P_{e} (\vec{r}) = \int_{-\infty}^0 d\epsilon' g_{\vec{r}}(\epsilon').
	\label{Prob2}	
\end{equation}
Note that the spatial average of this local density of states equals the total density of states, $\frac{1}{N} \sum_i g_{\vec{r}_i}(\epsilon) = g(\epsilon)$.

Our single assumption is that the local density of states is only restricted by the stability criterion of Eqn. (\ref{Stability}). This criterion requires that there cannot be particles at distance $\vec{r}$ with onsite energy in the range $-\frac{V}{|\vec{r}|} < \epsilon < 0$. Consequently, $g_{\vec{r}}(\epsilon)$ must be zero in this range. Outside this excluded region (i.e., at lower energies), we assume that the local density of states is equal to the total density of states. Equating Eqn. (\ref{Prob1}) with Eqn. (\ref{Prob2}) at commensurate sites, we thus find
\begin{equation}
\label{int-eq}
	\frac{1}{2} - 2 \widetilde{\Pi} (\vec{r}) = \int_{-\infty}^{-\frac{V}{|\vec{r}|}} d\epsilon' g (\epsilon'),
\end{equation}
where the stability constraint of  Eqn. (\ref{Stability}) sets the upper bound on the energy integration.

While our interest lies in disorder free Coulomb type systems, we remark that the above considerations can be applied {\it mutatis mutandis} for any translationally and rotationally invariant interaction 
$V_{ij} = f(|\vec{r}_{ij}|)$ (wherein Eqn. (\ref{Stability}) generalizes to $\Delta E = \epsilon_{j} - \epsilon_{i} -V_{ij} >0$ and, accordingly, the upper limit of the integral of Eqn. (\ref{int-eq}) is replaced by $(-f(|\vec{r}|))$.

Taking the derivative with respect to the distance $\vec{r}$ of both sides of Eqn. (\ref{int-eq}), we find at large distances
\begin{equation}
	\frac{1}{\sqrt{r}} e^{-r/\xi} \sim \frac{V}{r^2} g \left(- \frac{V}{|r|} \right).
\end{equation}
This implies that at low energies, the functional shape of the density of states should be
\begin{equation}
	g(\epsilon) \sim e^{-\frac{V}{ \xi |\epsilon|}},
	\label{BasicRelation}
\end{equation}
consistent with our numerical results, up to a possible power-law prefactor.


\section{Relation to quenched disorder systems}

The gap we have found numerically in our model is stronger, with a more suppressed density of states, compared to the usual Coulomb gap in systems with quenched disorder. There the Efros-Shklovskii bound is saturated $g(\epsilon) = C |\epsilon|^{d-1}$, with $C$ a universal disorder-independent pre-factor\cite{1975JPhC....8L..49E,1976JPhC....9.2021E,Shklovskii:1984wq}. Now in general, we know that for large disorder strength there are no charge correlations other than the correlation hole around $\vec{k}=0$ associated with the Coulomb tail. In this limit, the assumptions underlying Eqn. (\ref{int-eq}) are invalid, and consequently our analysis of the previous section cannot be used. In this regime the Efros-Shklovskii bound can be saturated. However, when disorder is weak compared to the Coulomb energy scale it is an open question how the gap changes from a power-law to an exponentially soft gap.

We verified this by introducing onsite disorder,
\begin{equation}
	E_{\mathrm{dis}} = \sum_i \mu_i n_i
\end{equation}
where $\mu_i$ is randomly chosen from the uniform distribution $[-W,W]$. Instead of averaging over the manifold of metastable states, we now average over different disorder realizations.



\begin{figure}
  \includegraphics[width=\columnwidth]{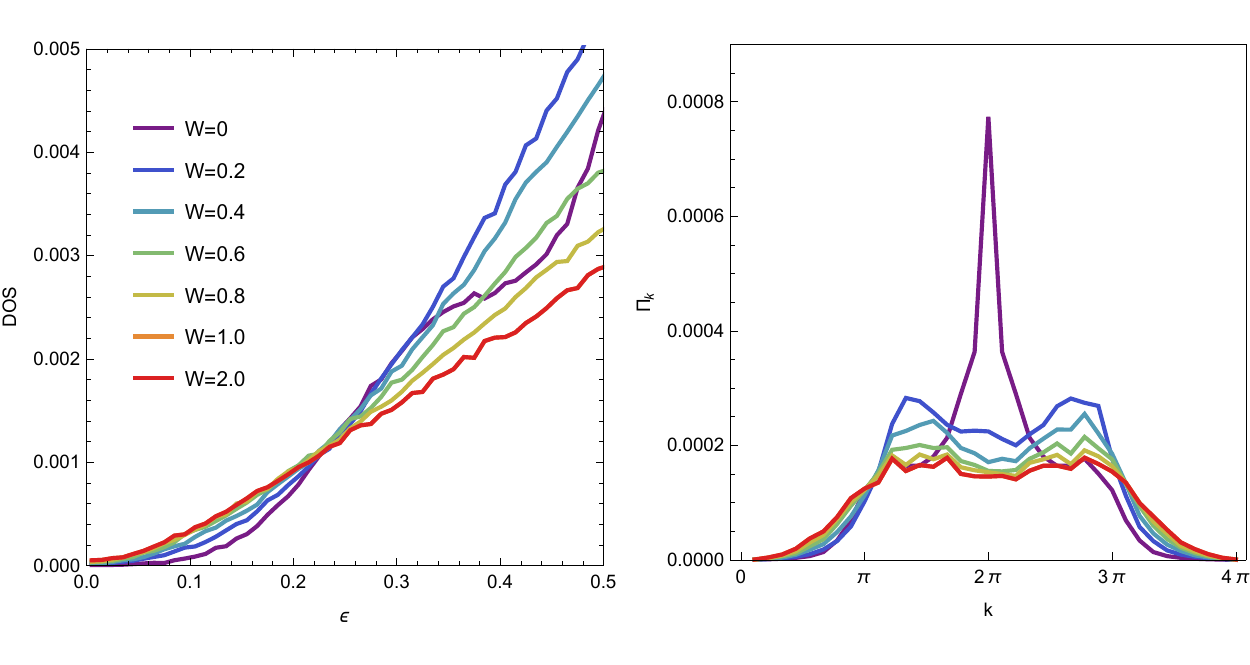}
  \caption{
  {\bf Left:} The density of states upon increasing the quenched disorder on a $L=36$ size lattice for 1000 disorder realizations. The density of states displays a smooth crossover from the exponential structure at $W=0$ to a power-law form. \\
  {\bf Right:} The density-density correlations along the line $(0,k)$ in the momentum space, for various disorder strengths (color coding is same as on the left side). Upon introduction of quenched disorder, we rapidly lose the peak at the $M$-point $(0, 2\pi)$ associated with local charge correlations. Only the broad feature at the edge of the Brillouin zone remains.}
  \label{disorderfig}
\end{figure}

In Fig. \ref{disorderfig} we show our results for 1000 disorder realizations for various disorder strengths. There seems to be a smooth transition from the exponential gap in the absence of disorder to a power-law gap. Also the charge correlations, crucial for our understanding of the exponential gap, seem to disappear upon inclusion of disorder. However, these results should be treated with caution. It is well-known that finite-size effects are increasingly important for systems with quenched disorder, when studied on a square lattice.\cite{Glatz08,Mobius} We therefore only present these results to show that the relation between quenched disorder and the exponential gap warrants further study.

\section{Outlook}
We proposed that self-generated glasses with Coulomb interactions have an exponentially suppressed density of states, following Eqn. (\ref{BasicRelation}). This relation is satisfied in a simple model of particles on a triangular lattice that was shown earlier to form a glass\cite{Mahmoudian:2014vh,Rademaker:2016fj}.

In self-generated electron glasses, our result can be directly studied by performing tunneling experiments. The structure of the low-energy density of states can be measured indirectly via the DC conductivity. For structural glasses composed of atoms or molecules it is difficult to measure the density of states directly. In general, the existence of single-particle excitations of arbitrary low energy lead to characteristic dynamical properties such as crackling and avalanches. However, because our model does not saturate the Efros-Shklovskii bound, we do not expect a scale-invariant avalanche distribution.\cite{Wyart:2012gz,DeGiuli:2014fb,Muller:2015kl,2015arXiv150103017Y}

We verified our hypothesis using a simple model of Coulomb interacting particles on a triangular lattice. It would be interesting to see whether models that exhibit icosahedral local order also display the exponentially soft gap.\cite{2013PhRvB..87r4203S} Furthermore, we expect a similar gap structure on other sufficiently frustrated lattices as long as there is a first-order transition that can be supercooled to form metastable states, such as in the 1/3-filled square lattice\cite{Rademaker2013} or the 3d pyrochlore lattice. The arguments presented in this Letter generalize trivially to an arbitrary power-law interaction of the form $V/r^\gamma$. However, interactions that decay faster than the dimensionality of the system, $V \sim 1/r^\gamma$ with $\gamma > d$ do not lead to glassy behavior. Indeed we were not able to reproduce a soft gap for the triangular lattice model with dipolar interactions. It remains an interesting open question whether long-range interactions are a sine qua non for glass formation.\cite{Berthier:2011fpa,Muller:2015kl}

Note that we only studied the density of single-particle excitations. Many-particle excitations, especially the ones that transition from one metastable state to another, play an important role in the slow dynamics of glasses. A full theory of glass-formation would treat both these many-particle events and the single-particle excitations at the same level. However, this is outside the scope of the current paper.

In this Letter, our considerations and analysis centered on the zero-temperature ensemble of metastable states. At finite temperature the gap will be filled, and earlier results are consistent with an exponentially weak scaling at finite temperature, $g(\epsilon=0, T) \sim T^{-1/2} \exp(-V/T)$ \cite{Rademaker:2016fj}. Again, notice the relative stability compared to systems with quenched disorder where $g(\epsilon=0,T) \sim T$.\cite{Menashe:2001fo,Somoza:2008dk} Finally, real glasses are obtained by a fast quench after which the gap needs time to develop. In fact, it has been shown that the soft gap forms extremely slowly, with power-law or even logarithmic time dependence $g(\epsilon=0,t) \sim (\log t)^\xi$ \cite{Huse:1986dx,Yu:1999jg,Grempel:2007kz,DiazSanchez:2001gi}. In our case, however, the absence of marginal stability opens up the possibility of a true thermodynamic phase transition into a glass phase. How the glass, with its soft gap and the concomitant local density correlations, may be dynamically generated in various systems following a quench to nonzero temperatures is an interesting question for future research.

\ack We are thankful to J.~Zaanen, J.~S.~Langer, M. Ortu\~{n}o, L.~Cugliandolo and L. Yan for discussions. We especially acknowledge G. Parisi for his insights. L.R. was supported by the Dutch Science Foundation (NWO) through a Rubicon grant. Z.N. acknowledges support by the NSF under grant no. DMR-1411229 and the Feinberg Foundation for visiting faculty at the Weizmann Institute. 
L.B. was supported by the NSF Materials Theory program, grant number DMR-15-06119.
V.D. was supported by the NSF grants DMR-1005751 and DMR-1410132.

\end{document}